\documentclass[fleqn,usenatbib]{mnras}

\usepackage{newtxtext,newtxmath}

\usepackage[T1]{fontenc}

\DeclareRobustCommand{\VAN}[3]{#2}
\let\VANthebibliography\thebibliography
\def\thebibliography{\DeclareRobustCommand{\VAN}[3]{##3}\VANthebibliography}

\usepackage{graphicx}	
\usepackage{amsmath}	
\usepackage{graphicx}
\usepackage{epstopdf}  

\title[The central region of quasars]{Research on the central region of quasars based on variability and structure function}

\author[X.Wei et al.]{
X. Wei,$^{1,\dagger }$
J. Tang,$^{1,\dagger}$\thanks{E-mail: tj168@163.com}
Y. Tao$^{1}$
and X. H. Zhang$^{1}$
\\
$^{1}$School of Physics and Telecommunication Engineering, Shaanxi University of Technology, Hanzhong 723000, China\\
}

\date{Accepted XXX. Received YYY; in original form ZZZ}

\pubyear{\the\year{}}

\begin{document}
\label{firstpage}
\pagerange{\pageref{firstpage}--\pageref{lastpage}}
\maketitle

\begin{abstract}
Quasars, as extremely luminous and distant special celestial bodies in the universe, are driven by a complex system composed of supermassive black holes and surrounding accretion disks. This paper adopts a time-domain observation strategy and combines the analysis of light curves with the construction of structure functions to indirectly reveal the physical essence of the central region of quasars from the perspective of variability. The research data are derived from the large sample observation data of the Sloan Digital Sky Survey (SDSS). Through extensive data statistics and correlation analysis, a series of important findings have been obtained: the characteristic parameters of the structure function of quasars show significant correlations with luminosity, black hole mass, and Eddington ratio. That is, quasars with higher luminosity, larger black hole mass, and larger Eddington ratio have larger structure functions. For quasars of the same luminosity, the larger the Eddington ratio, the smaller the structure function. However, the correlation between the structure function and redshift or rest wavelength is not significant, indicating that the variability characteristics of quasars are mainly determined by their own physical properties and are minimally affected by the cosmological redshift effect.
\end{abstract}




\section{Introduction}

Active Galactic Nucleus (AGN) is a very special component of extragalactic galaxies. Within an AGN, there are intense activities and physical processes that occur. This makes it one of the most highly studied subjects in astrophysics. AGNs can be classified into multiple subcategories based on different characteristics, mainly including quasars, quasar-like objects, Seyfert galaxies, etc. Quasars (QSO) are among the brightest AGNs in the universe. Their core feature lies in the extreme luminosity they possess. Typical quasars can emit radiation with a luminosity tens of trillions of times that of the Sun in the optical band, and the energy comes from the intense accretion process of the supermassive black hole (SMBH) at the center onto surrounding matter \citep{Czerny2023, Marziani2025}. The existence of this accretion process efficiently converts gravitational potential energy into radiation, thereby driving the extremely high luminosity output of quasars.

The continuous advancement of technical means has expanded the originally scarce astronomical database, and the research content on quasar variability has also been continuously enriched. Based on the continuously enriched variability database, previous studies have conducted multi-angle research on the nature of quasars and the underlying physical mechanisms.

\citet{Aretxaga1997} pointed out that by performing exponential fitting based on the typical structure function, the variability timescale can be estimated. \citet{Cristiani1996} used 180 quasars as a sample to study the correlation between the variability timescale and absolute magnitude and redshift, and concluded that there was no significant correlation. In addition, to explore the variability mechanism of AGNs, many researchers have analyzed the dependence of variability indicators on luminosity, black hole mass, and rest-frame wavelength. However, the sample size was relatively small, and other parameters also affected the dependence results. Regarding the dependence of variability indicators on the rest-frame wavelength, there is an inverse correlation between wavelength and variability \citep{Cutri1985, Paltani1994, diClemente1996}, and \citet{Berk2004} verified this inverse correlation after excluding the influence of luminosity and redshift.

For the dependence on other parameters, there are similar issues, due to the incompleteness of small samples, leading to controversial conclusions \citep{Bonoli1979, Netzer1983, Cristiani1990, Giallongo1991, Hook1994}. Although later studies used large samples to verify the dependence on other parameters \citep{Wilhite2008, Bauer2009}, using the variability indicators to study led to the neglect of the diversity of individual sample variability curves, which may lead to deviations from the actual results. Currently, multiple studies have been conducted on the dependence of variability and redshift, luminosity, rest-frame wavelength, Eddington ratio, and black hole mass, but the sample sizes are too small and the influence of parameters is not excluded \citep{Wold2007, Ai2010, Tang2023}.

In the research of Zuo Wenwen \citep{Zuo2012}, the variability amplitude was used to quantify the variability, and the dependence of this indicator on luminosity, redshift, Eddington ratio, and black hole mass was explored. It was pointed out that the luminosity amplitude was negatively correlated with the rest-frame wavelength, and showed a significant negative correlation with the Eddington ratio and luminosity, while the dependence on black hole mass varied under different parameter control conditions. In the Eddington ratio bins, the variability amplitude was negatively correlated with black hole mass; in the luminosity bins, the variability amplitude was positively correlated with black hole mass. 

In Zuo Wenwen's research, the binning thinking and individual samples were taken into account, avoiding the limitations caused by considering the collective indicators. With the continuous advancement of technical means and the continuous update of equipment, the variability curve database is becoming increasingly abundant, meeting the demand for a large number and high-quality samples for research. Therefore, there has been more comprehensive research on the dependence relationships of various parameters. \citet{Yu2025} used the four key parameters fitted by the DHO model (improved DRW model) to study the correlation with quasars' Eddington ratio, rest-frame wavelength, and black hole mass. \citet{Goncalves2025} conducted binning operations on the redshift parameters to explore the universal relationship between the amplitude of photometric variations and the Eddington ratio within different redshift ranges. \citet{Jha2025} used two indicators obtained from the DRW model fitting to investigate the positive correlation between the optical variability of Type 1 active galactic nuclei and the black hole mass luminosity, and analyzed the correlation between optical variability and radio and X-ray properties, distinguishing the differences in the variability mechanisms across three bands. \citet{Burke2020} analyzed the correlation between the characteristic times obtained from the DRW model fitting and the mass of the supermassive black hole, finding that the mass of the supermassive black hole has an impact on the variability of the active galactic nucleus. \citet{Wang2023}  discovered a dependency relationship between the luminosity, the mass of the supermassive black hole, and the amplitude of the variability. Based on the DRW model, \citet{Chen2023} discovered a dependency between the mass of the supermassive black hole and the damping timescale through the variability in the infrared band. As a common method for studying variability, the study of the structure function naturally plays a crucial role. The study of the structure function is used to infer possible variability mechanisms \citep{Hawkins2002}. \citet{Kozlowski2016} used the structure function to study the variability of quasars in SDSS, and based on the covariance function of random processes, studied the correlation of the physical parameters of quasars. Although recent studies still focus on the dependency relationship between variability indicators and physical parameters, there is still no clear conclusion \citep{Sanchez2018, Sanchez2019, Lu2019, Xin2020, Suberlak2021, deCicco2022, Stone2022, Arevalo2023, Arevalo2024}.

In this study, we adopted the binning thinking and calculated the correlation coefficients between the structure function of a single sample and different physical parameters. According to the correlations within different parameter ranges, we explored the variability mechanism. In the second section, we introduced the research samples and methods used in this study. In the third section, we presented the parameter binning of the samples and the dependencies between the structure function and various physical parameters. In the fourth section, we analyzed and discussed the results of the study. In the fifth section, we presented our conclusions.

\section{Samples and methods}

In this study, the sample of quasars with photometric variability consists of 6571 objects, which were obtained through matching with SDSS \citep{Schneider2010, Shen2011}. It includes data from five bands (u, r, i, g, z) over a period of more than ten years. These observational data are used to calculate the structure function of the long-term photometric variability of quasars. In addition to the photometric data, physical parameters of each sample are also required in the subsequent correlation calculations. The physical parameters obtained by reading previous research results include black hole mass, luminosity, and Eddington ratio \citep{Shen2011}. During the preprocessing of the photometric data, the redshift of the samples is obtained. Besides, we calculated the accretion rate using the obtained parameters \citep{Shakura1973, Merloni2003}.

\label{sec:maths} 
\begin{equation}
    \dot{M} = 1.13 \times 10^{-8} \lambda_{\rm Edd} \left( \frac{M_{\rm BH}}{M_\odot} \right).
	\label{eq:quadratic}
\end{equation}

During the calculation of the accretion rate, the standard thin disk assumption was adopted, and the radiation efficiency was set as $\eta$ = 0.1 \citep{Einstein1936}.

\subsection{Samples}

In this study, the photometric data used for the correlation analysis were sourced from the SDSS survey. The Sloan Digital Sky Survey (SDSS) employs a 2.5-meter telescope at the Apache Point Observatory in the United States, providing photometric data and spectral data across five bands. The large sample size is a typical feature of the SDSS survey. By 2014, the SDSS survey had discovered approximately 166,583 quasars \citep{Paris2012}, and the number of observable quasars has now reached tens of thousands. Since its development, the SDSS survey has undergone multiple upgrades, with continuous improvements in observation accuracy and an increasing observation duration \citep{Tang2024, Krolik1999}. Moreover, the scientific goals of the SDSS survey cover various cutting-edge fields such as galaxy evolution, large-scale structure of the universe, quasar detection, and stellar physics \citep{Tang2023}. As of 2024, the SDSS survey has released version DR18, which includes over 500,000 spectral identification results of quasars and multi-band photometric data. The massive data volume and the principle of open sharing have made the SDSS one of the most core data sources for quasar photometric variability studies \citep{Tang2024, Manmoto1996}.

\subsection{Methods}

\subsubsection{Structure function}

As a common method for studying the variability of galactic nuclei, the structure function is used to analyze the variability characteristics of celestial bodies. The advantage of the structure function lies in avoiding the window problem and aliasing phenomenon in Fourier analysis while being able to analyze data sequences with sparse and uneven data points. Through the structure function, the root mean square value of the difference in variability intensity corresponding to different times in the variability curve can be represented \citep{Simonetti1985}. The nearly linear part in the typical graph of the structure function over time delay has a physical meaning related to the rate of change of flow \citep{Tang2012}. The slope of this part can be used to analyze the intrinsic variability of active galactic nuclei and quasars, and to some extent, can reveal the physical mechanism behind the variability \citep{Hawkins1996, Hawkins1993}. Wang Hongtao \citep{Wang2020} used the structure function method to study the variability characteristics in the infrared band of quasars in the SDSS survey.

For the observed value $X(i)$ of a certain celestial body, taking any time delay $\tau$, its first-order structure function $SF(\tau)$ is as shown \citep{Simonetti1985, Butuzova2025}

\label{sec:maths} 
\begin{equation}
    SF(\tau)=\frac{1}{N(\tau)}\sum_{i}^{N}\left[X(i+\tau)-X(\tau)\right]^2
	\label{eq:quadratic}
\end{equation}
where $X(i)$ and $X(i+\tau)$ represent the observed value of the $i$ and $i+\tau$ points, respectively, $N(\tau)=\sum [\omega(i)\omega(i+\tau)]$represents the number of the observations. If observed values exist at the $i$ and $i+\tau$ points, the weight coefficient is $\omega(i)=\omega(i+\tau)=1$, and if not, $\omega(i)=\omega(i+\tau)=0$.

\subsubsection{Correlation analysis}

The Spearman correlation coefficient (Spearman) is used to assess the strength and direction of the monotonic relationship between two variables. Since this correlation index does not emphasize the normal distribution of data and does not have mandatory requirements for the linear relationship between variables, it can better reflect the correlation in the analysis of the correlation between ordered data \citep{Spearman1904}. In this paper, r1 is used to represent the Spearman correlation coefficient between the structure function and each physical parameter.

\label{sec:maths} 
\begin{equation}
    r_1 = 1 - \frac{6\Sigma d_i^2}{n(n^2 - 1)}.
	\label{eq:quadratic}
\end{equation}
In the formula, $di$ represents the difference in ranks between the two variables for each observation pair, and $n$ is the sample size.

The Pearson correlation coefficient (Pearson) is the most widely used measure of correlation in research, commonly used to quantify the strength and direction of the linear relationship between two continuous variables. When calculating the Pearson correlation coefficient, it is assumed that the data are continuous and follow a normal distribution, and it is required that there is a linear relationship between the variables \citep{Pearson1895}. Compared with the Spearman correlation coefficient, if there is no obvious linear relationship between the data, the calculated Pearson correlation coefficient will be significantly smaller than the Spearman correlation coefficient. In this study, r2 is used to represent the Pearson correlation coefficient between the structural function and each physical parameter.

\label{sec:maths} 
\begin{equation}
    r_2 = \frac{\sum_{i=1}^n (x_i - \bar{x})(y_i - \bar{y})}{\sqrt{\sum_{i=1}^n (x_i - \bar{x})^2} \sqrt{\sum_{i=1}^n (y_i - \bar{y})^2}}.
	\label{eq:quadratic}
\end{equation}
In the formula, $xi$ and $yi$ represent the values of individual sample points, $\bar{x}$ and $\bar{y}$ represent the sample means of the corresponding variables, and $n$ represents the sample size.

Whether it is the Spearman correlation coefficient or the Pearson correlation coefficient, their numerical range is between -1 and 1. 0 indicates no correlation, 1 indicates a perfect positive correlation, and -1 indicates a perfect negative linear correlation. The larger the absolute value of the two correlation coefficients, the stronger the correlation they represent. Conversely, the smaller the absolute value, the weaker the correlation.

\section{Correlation analysis}

This chapter is based on the SDSS large sample of quasar observations. It constructs the distribution of four physical parameters (luminosity, redshift, black hole mass, and Eddington ratio) for 6571 selected samples. On this basis, the correlation coefficients are calculated in bins. Since the black hole mass, Eddington ratio, and width have mutual influences, when studying the correlation between the structure function and one of these parameters, the influence of the other two parameters may be affected. To minimize the impact of other parameters on the research results, when studying the correlation between the structure function and a certain parameter, the other parameters are restricted within a certain range, and the correlation between the structure function of the sub-sample set and other parameters is explored within this range. Based on the given parameters, the correlation results between the structure functions of several sub-sample sets and the parameters are used to analyze the dependency relationship between the structure function and parameters of the entire data set.

\subsection{The dependence of the structure function on redshift}

When exploring the correlation between the sample structure function and redshift, the ranges of the Eddington ratio, black hole mass, and rest wavelength were restricted. The original large sample set was divided into 25 sub-sample sets, and the Spearman correlation coefficient r1 and Pearson correlation coefficient r2 of the structure function and redshift were calculated in each sub-sample set. Linear equation fitting was performed within the allowable error range. The specific ranges of the binning parameters and the correlation coefficients and fitting slopes were marked in each sub-sample set.

In Figure 1, the gray scattered points represent the original data points. Based on the density of these scattered points, it can be found that the sample abundance in the sub-sample sets is sufficient. In most sub-plots, the scattered points show a continuous distribution along the redshift. The samples are most dense in the range of redshift from 0.5 to 2.0, indicating that the selected samples are mainly concentrated in the low and medium redshift regions. The red points are the fitting points, representing the statistical mean after binning the structure function in different redshift intervals. The fitting curves formed by these fitting points can to some extent reflect the overall evolution trend of the structure function and redshift.

Overall, the correlation between the structure function and redshift is weak in all sub-plots. Most samples and redshift show a weak negative correlation, and a few samples show a positive correlation. The minimum value of r1 is -0.177, and the maximum value is 0.176, showing a weak correlation; the minimum value of r2 is -0.195, and the maximum value is 0.182. From the results of Spearman correlation coefficient and Pearson correlation coefficient, the correlation between the structure function of quasars and redshift is not significant.

Among these sub-sample sets, both high Eddington ratios and low Eddington ratios show a weak negative correlation. In the sub-plots where the Eddington ratio is less than -0.82, nearly 60\% show a trend that the structure function decreases slowly with the increase of redshift, and the remaining sub-plots have almost no correlation; in the sub-plots where the Eddington ratio is greater than -0.82, approximately 70\% of the samples have a weak negative correlation with redshift, and the correlation in the remaining sub-plots is extremely weak. From the grouping of Eddington ratios, the samples with negative correlation in high Eddington ratios are higher than those in low Eddington ratios. In addition to the Spearman correlation coefficient and Pearson correlation coefficient, the fitting lines also show a downward trend, verifying the weak negative correlation trend.

\begin{figure*}
	\includegraphics[width=\textwidth]{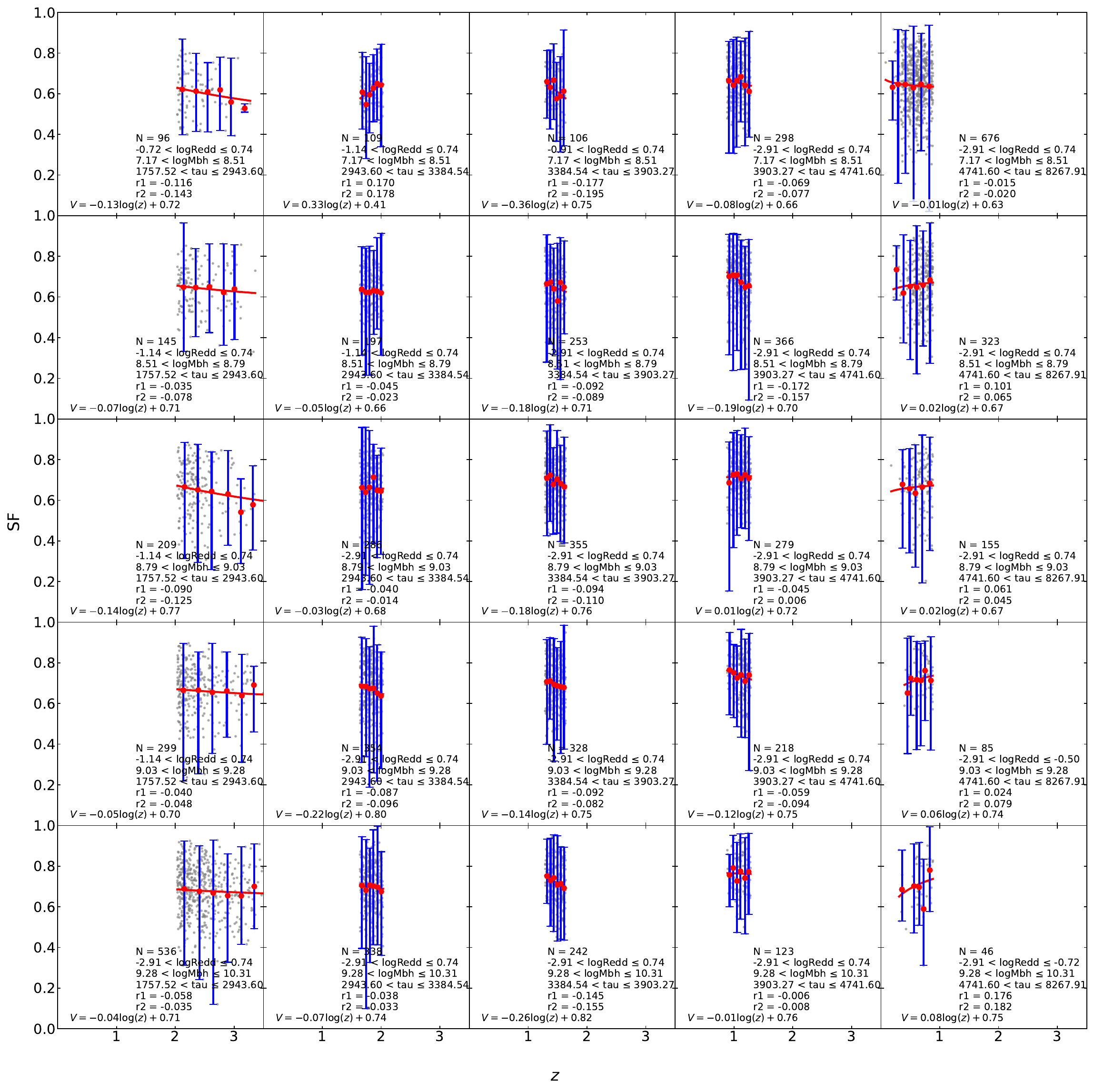}
    \caption{The dependence of the structure function on the redshift z. In the figure, restrictions are imposed on the black hole mass, the Eddington ratio, and the stationary wavelength, and 25 sub-sample sets are set up.}
    \label{fig:example_figure}
\end{figure*} 

\subsection{The dependence of the structure function on the stationary wavelength}

In the study of the structure function of quasars and the dependence on the stationary wavelength, the sub-sample sets were binned based on the ranges of redshift, Eddington ratio, and black hole mass. Similarly, 25 sub-sample sets were divided. The dotted lines and annotations in the figure are the same as those in Figure 1.

From the number of samples in each sub-sample set in Figure 2, it can be seen that the overall sample is sufficient, and the sample distribution within each wavelength interval is continuous without obvious breaks, proving that the sub-sample sets after binning can achieve basic coverage in the wavelength dimension. It is easy to notice that in most sub-plots, there is a slow upward trend of the structure function with the stationary wavelength. The maximum value of r1 is 0.230 and the minimum value is -0.228, the maximum value of r2 is 0.255 and the minimum value is -0.212. However, overall, there are fewer sub-plots showing a negative correlation, and most sub-plots show a weak positive correlation. In the short wavelength range, the structure function increases rapidly, while in the medium and long wavelength range, the growth of the structure function gradually becomes more gradual. This indicates that in the short wavelength interval, the structure function of quasars is more sensitive to wavelength changes, while in the medium and long wavelength range, the structure function tends to be stable.

Since there is also a grouping of Eddington ratio in Figure 3, it can be seen from the figure that in the sub-plots of the low Eddington ratio group, the correlation coefficient between the structure function and the wavelength is relatively lower than that of the high Eddington ratio group, proving that the Eddington ratio not only has a modulation effect in the dependence on redshift but also has a similar modulation mechanism in the dependence on wavelength.

\begin{figure*}
	\includegraphics[width=\textwidth]{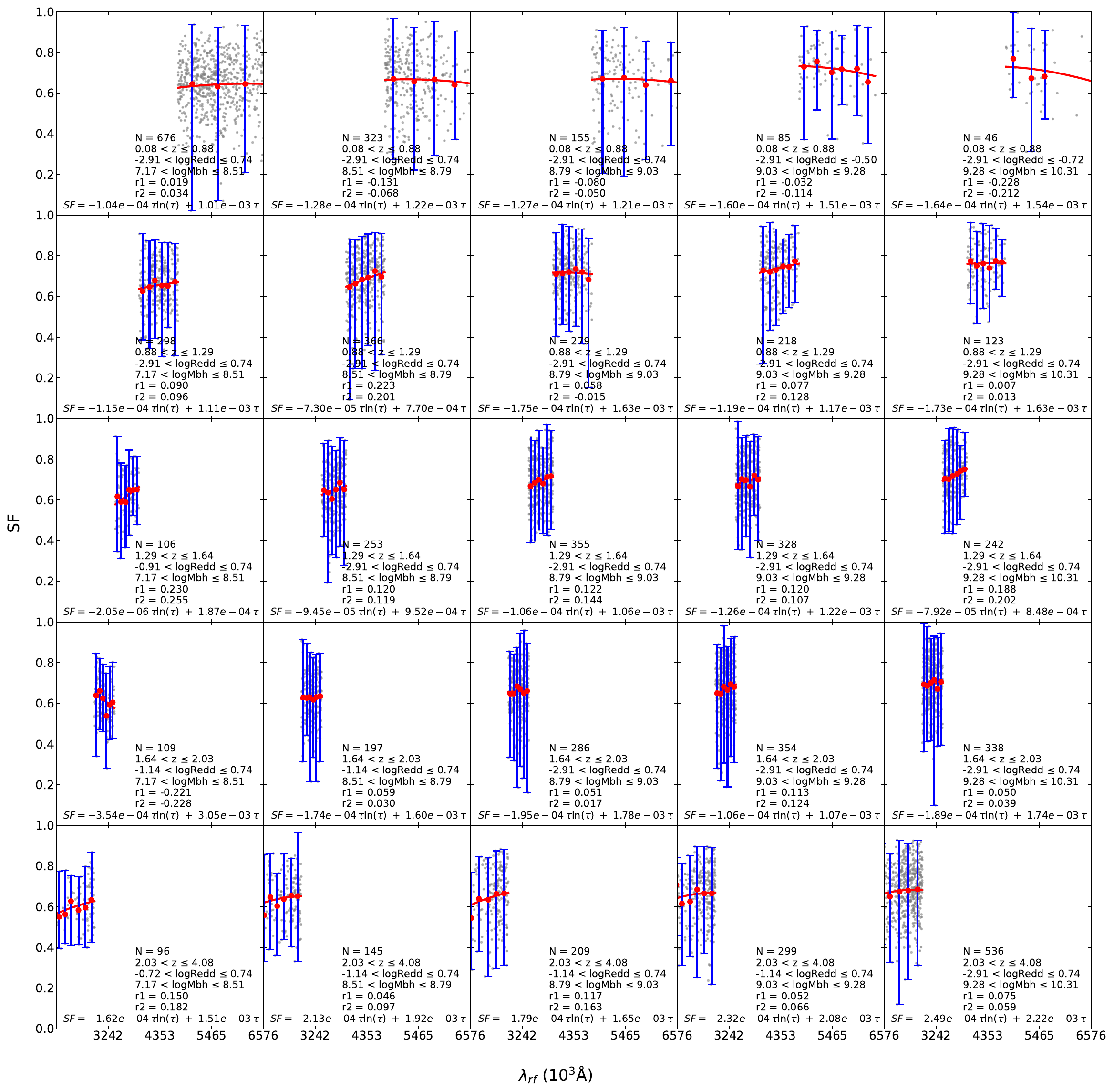}
    \caption{The dependence of the structure function on the rest-frame wavelength. In the figure, the black hole mass, Eddington ratio, and redshift are restricted, and 25 sub-sample sets are set.}
    \label{fig:example_figure}
\end{figure*} 

\subsection{The dependence of the structure function on luminosity}

Firstly, we explore the dependency relationship between the structural function and luminosity within the Eddington ratio range limit. As indicated in the previous figure, in Figure 3, the samples, correlation coefficients, and fitting equations in the sub-sample sets have all been presented.

From Figure 3, it can be seen that the sample quantities within each sub-sample set are sufficient, basically avoiding the bias caused by small samples, proving the reliability and rationality of binning. The sample points show that the luminosity presents a continuous and dispersed distribution within the range of 45.0 - 48.0, with no obvious interval discontinuity, indicating that each group covers the light intensity dimension completely. From the trend of the fitting line, the structural function of quasars and luminosity show a positive correlation overall, which contrasts with the weak correlation characteristics of the structural function with redshift and static wavelength previously, indicating that the modulation effect of luminosity on the light variation of quasars is more significant. The subplots can be divided from left to right into low redshift group and high redshift group. In the low redshift group, the trend of the structural function changing with luminosity is relatively gentle, while in the high redshift group, the trend is more steep. Besides the redshift grouping, the correlation and redshift of the structural function with luminosity in the Eddington ratio grouping situation also show a similar trend. In the high Eddington ratio grouping, the correlation coefficient is larger, while in the low Eddington ratio grouping, the correlation coefficient significantly decreases. In these sub-sample sets, the maximum value of r1 is 0.439 and the minimum value is 0.091; the maximum value of r2 is 0.433 and the minimum value is 0.102. Whether it is the Spearman correlation coefficient or the Pearson correlation coefficient, the presented correlation is much larger than that of the structural function with redshift and the structural function with static wavelength. It can be speculated that the luminosity has a certain influence on the light variation of quasars, and the influence degree is more intense than redshift and static wavelength.

The analysis shows that the higher the Eddington ratio, the more obvious the positive correlation tendency of the structural function of quasars increasing with luminosity; the lower the Eddington ratio, the positive correlation tendency of the structural function with luminosity is slightly weaker, and the proportion of negative correlation samples is slightly higher. This feature is in synergy with the conclusion in the previous high redshift grouping that the correlation between the structural function and luminosity is stronger, suggesting that high Eddington ratio and high redshift quasars share common characteristics in accretion physics.

\begin{figure*}
	\includegraphics[width=\textwidth]{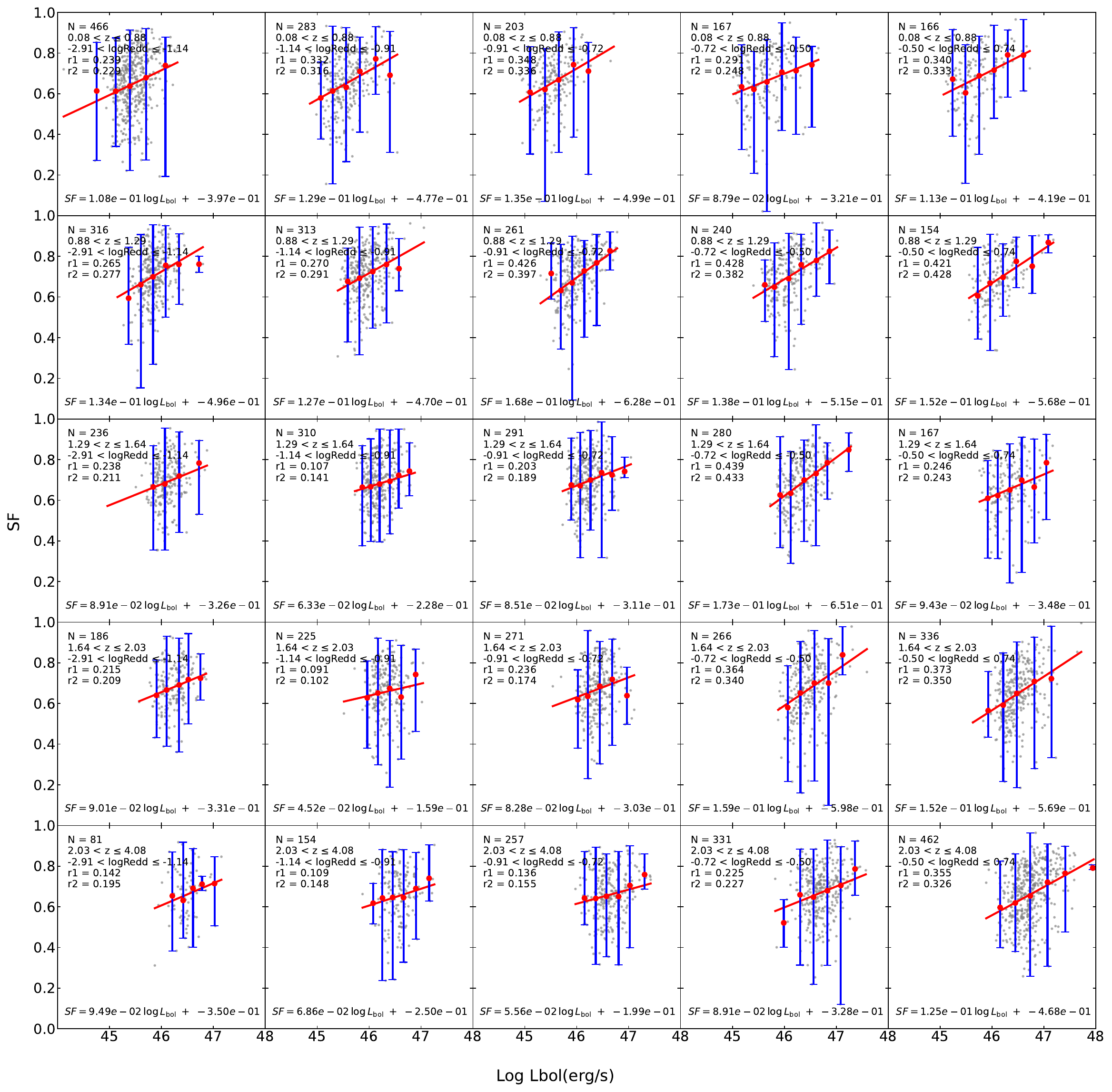}
    \caption{The dependence of the structural function on luminosity. In the figure, restrictions were imposed on the Eddington ratio and redshift, and 25 sub-samples were set.}
    \label{fig:example_figure}
\end{figure*} 

Secondly, the dependence relationship between the structure function and luminosity within the mass range limit of black holes was explored. The annotations in the previous figure remain the same. In the figure, the samples, correlation coefficients, and fitting equations in the sub-sample sets are all presented.

According to Figure 4, the sample quantities within each sub-sample set are sufficient. When dividing the black hole mass into bins, small sample sizes did not cause deviations, proving the reliability of binning. The sample points show a continuous and dispersed distribution throughout the luminosity range, with no obvious interval discontinuities, indicating that each group covers the entire luminosity dimension. From the trend of the fitting line, the structure function and luminosity of quasars in the limit of black hole mass also show a positive correlation as a whole. This is in contrast to the weak correlation characteristics of the structure function with redshift and the structure function with stationary wavelength previously observed. Combined with the binning situation of the Eddington ratio, it indicates that the modulation effect of luminosity on the light variation of quasars is more significant. In these sub-sample sets, the maximum value of r1 is 0.416, and the minimum value is -0.035; the maximum value of r2 is 0.413, and the minimum value is -0.037. Whether it is the Spearman correlation coefficient or the Pearson correlation coefficient, the correlation presented is much larger than that of the structure function with redshift and the structure function with stationary wavelength. Although some sub-sample sets show a negative correlation, the proportion is very small, and the negative correlation is extremely weak. It can be speculated that the influence of luminosity on the light variation of quasars is more intense than that of redshift and stationary wavelength.

\begin{figure*}
	\includegraphics[width=\textwidth]{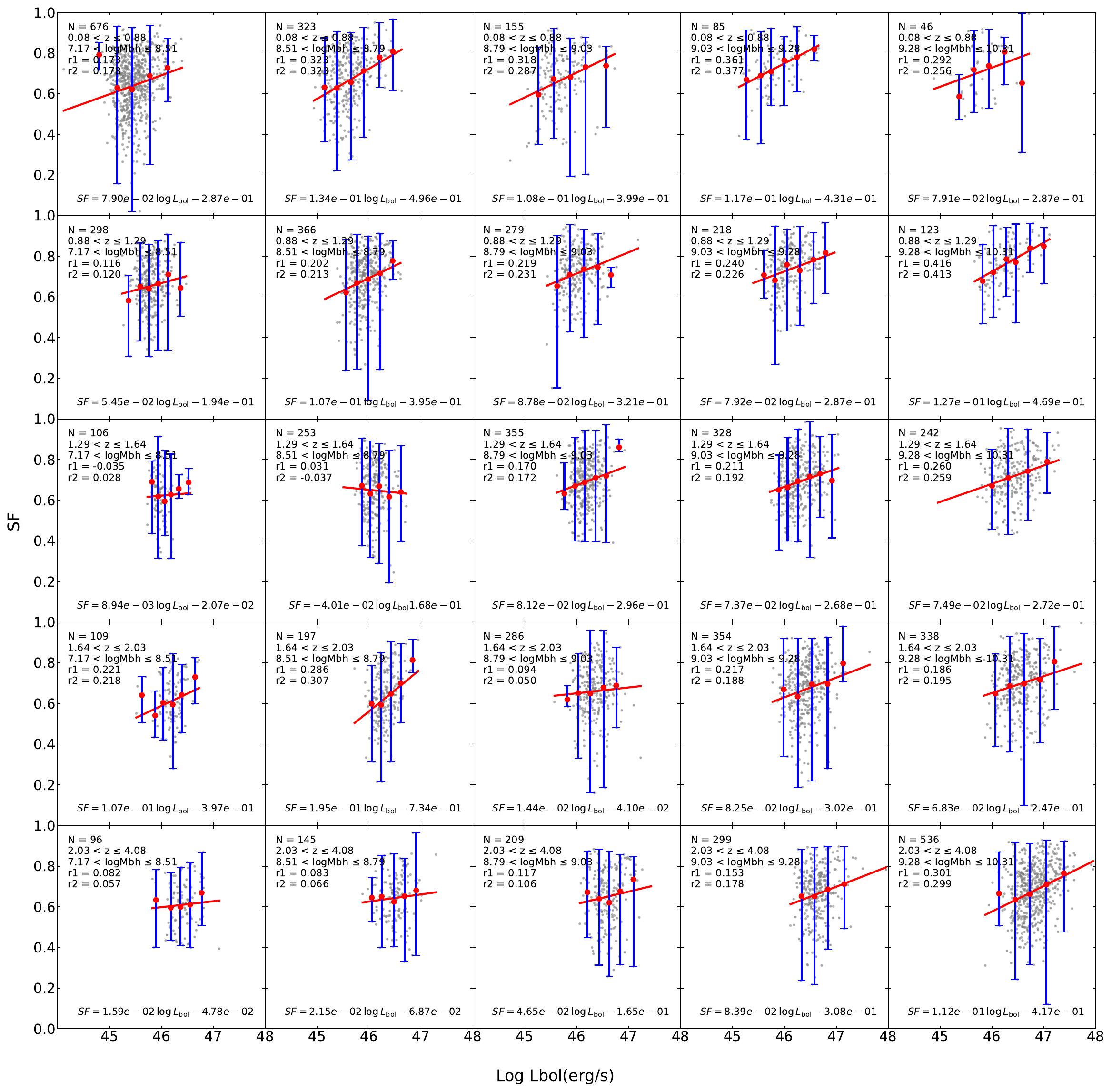}
    \caption{The dependence of the structure function on luminosity. In the figure, limits are imposed on the mass of the black hole and the redshift, and 25 sub-samples are set.}
    \label{fig:example_figure}
\end{figure*} 

\subsection{The dependence of the structure function on the mass of the black hole}

Firstly, we explore the dependency relationship between the structure function and the black hole mass within the luminosity range limit, with the same annotations as in the previous figure. In the figure, the samples, correlation coefficients, and fitting equations of the sub-sample sets are all presented.

According to Figure 5, the number of samples within each sub-sample set varies significantly under the binning of black hole mass, but the majority of the sub-sample sets have more than 100 samples, and the statistical results are relatively reliable. The distribution of black hole mass in the range of 7.5 - 10.0 is continuous, and there is no obvious interval break, which also indicates that the sample coverage in the black hole mass dimension is complete. From the correlation coefficient, the correlations of the sub-sample sets show both positive and negative correlations, but overall they present a positive correlation. The subgraphs from left to right correspond to the redshift and luminosity from low values to high values. It can be seen that in the sub-sample sets with relatively small dispersion, the correlation coefficient increases from left to right. In these sub-sample sets, the maximum value of r1 is 0.567, and the minimum value is -0.470. Considering that some sub-sample sets have a small number of samples and a large data dispersion, the data is re-estimated, and the maximum value is 0.497, and the minimum value is 0.007; the maximum value of r2 is 0.724, and the minimum value is -0.281. Similarly, the data is re-estimated, and the maximum value of the coefficient for the sub-sample set with a larger sample size is 0.385, and the minimum value is 0.014. However, whether it is Spearman's correlation coefficient or Pearson's correlation coefficient, the correlation presented is much larger than that of the structure function with redshift and the structure function with the stationary wavelength. It can be inferred that the influence of black hole mass on the light variation of quasars is more intense than that of redshift and the stationary wavelength.

\begin{figure*} 
    \includegraphics[width=\textwidth]{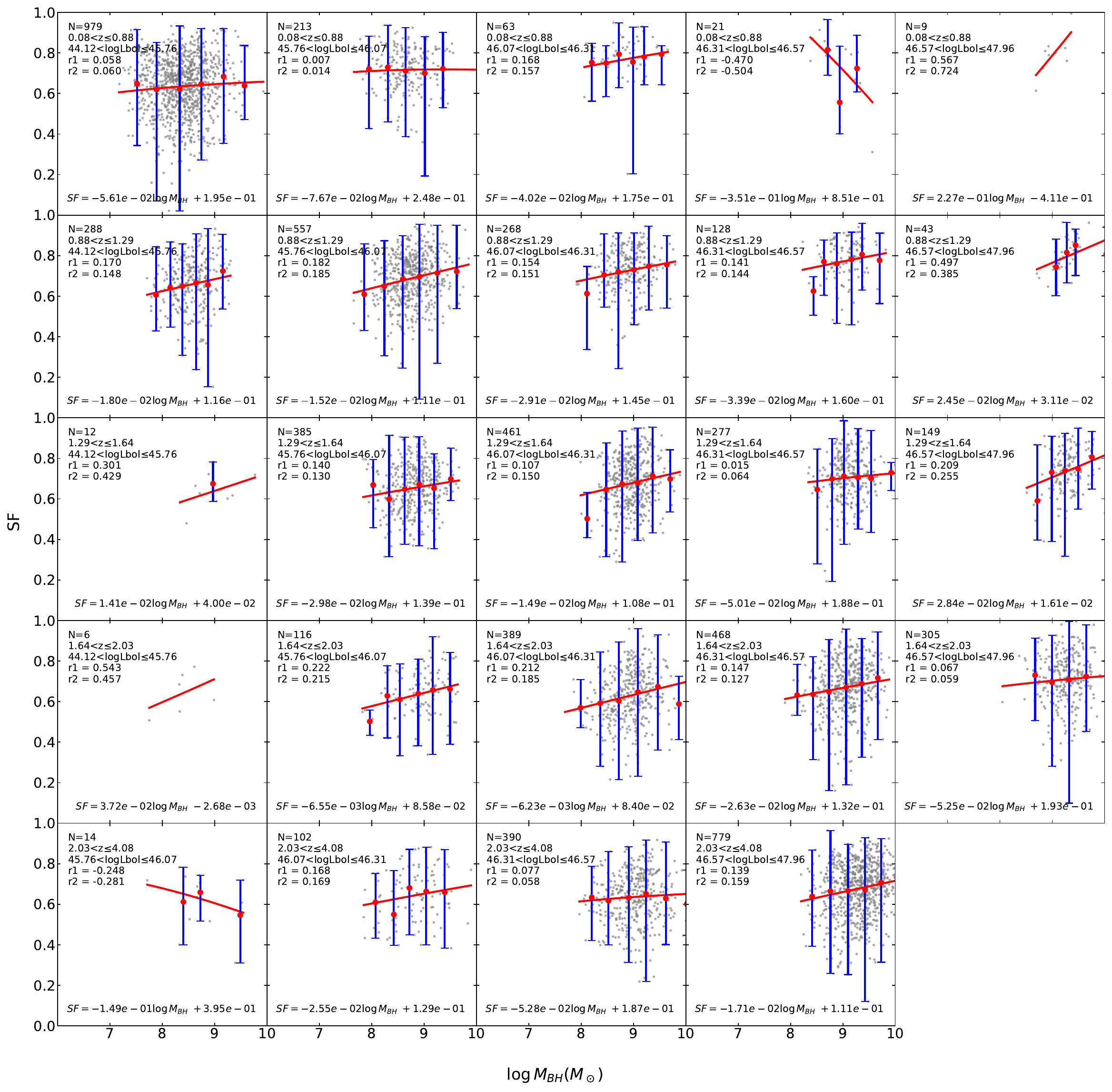}
    \caption{The dependence of the structure function on the mass of the black hole. In the figure, limits have been set on luminosity and redshift, and 24 sub-samples have been established.}
    \label{fig:example_figure}
\end{figure*} 

After completing the correlation coefficient statistical analysis under the photometric binning, the range of the Eddington ratio was restricted. Then, the dependence relationship between the structural function and the black hole mass was analyzed again, with the same annotations as in the previous figure. In the figure, the samples, correlation coefficients, and fitting equations in the sub-sample sets were all displayed.

Figure 6 shows the correlation coefficient situation in the sub-sample sets after binning. Overall, the sample quantity of each sub-sample set is slightly better than that in the photometric grouping situation, thus avoiding systematic errors caused by the discreteness of sample data. Similar to the photometric binning situation, in the black hole mass range of 7.5 - 10.0, the data points are continuously dispersed, proving the completeness of the black hole mass dimension. No obvious linear aggregation band is observed, and to a certain extent, it also indicates that the structural function is not determined by a single parameter but is influenced by multiple factors.

It is easy to notice that in the low redshift - low Eddington ratio grouping, the trend of the bin mean value is more gentle, indicating that the quasars in this parameter range evolve more slowly and the light variation is relatively gentle. While in the high redshift - high Eddington ratio grouping, the trend of the bin mean value is more steep, indicating that the accretion process of quasars in this range is more intense, and the structural function is more sensitive to the change of black hole mass, and the light variation is relatively intense. From the correlation coefficient perspective, the correlation of the sub-sample sets is positive correlation. The sub-figure from left to right corresponds to the situation of redshift and Eddington ratio from low value to high value. It can be seen that in these sub-sample sets, the maximum value of r1 is 0.436, the minimum value is 0.070, the maximum value of r2 is 0.443, and the minimum value is 0.087. However, whether it is the Spearman correlation coefficient or the Pearson correlation coefficient, the correlation presented is much larger than that of the structural function with redshift and the structural function with the stationary wavelength. Based on this analysis result, it can also be speculated that the influence of black hole mass on the light variation of quasars is more intense than redshift and the stationary wavelength.

\begin{figure*} 
	\includegraphics[width=\textwidth]{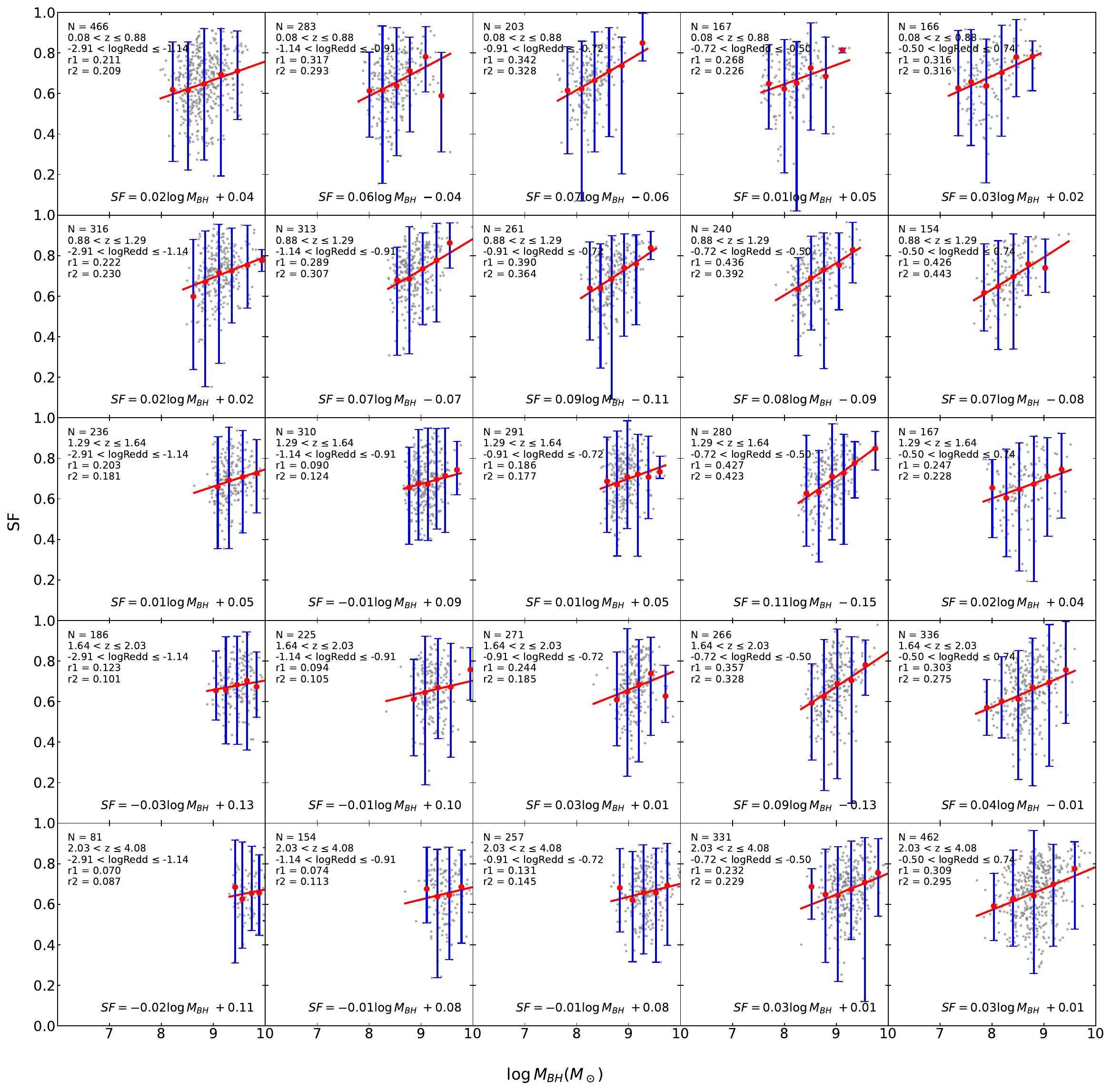}
    \caption{The dependence of the structure function on the mass of the black hole. In the figure, restrictions were imposed on the Eddington ratio and redshift, and 25 sub-samples were set.}
    \label{fig:example_figure}
\end{figure*} 

\subsection{The dependence of the structural function on the Eddington ratio}

Following the research methods presented in the previous section, the luminosity range was first restricted to explore the dependence relationship between the structural function and the Eddington ratio within this range. The annotations of the resulting graph are the same as those of the previous graph. The sample quantity, parameter range, and correlation coefficient are all labeled in the subgraphs.

As shown in Figure 7, the sample quantities within each sub-sample set within the luminosity bins vary significantly, but the majority of the sub-sample sets have more than 100 samples, and the statistical results are relatively reliable. The distribution within the Eddington ratio range of -2.05 to 0.85 is continuous, without any obvious interval fractures, which also indicates that the sample coverage in the Eddington ratio dimension is complete, consistent with the distribution characteristics of the 3-15 Eddington ratio frequency histogram. From the correlation coefficient perspective, the correlations within the sub-sample sets are positive and negative, but overall show a negative correlation. The subgraphs are arranged from left to right, corresponding to the redshift and luminosity from low values to high values. It can be seen that in the relatively less discrete sub-sample sets, the absolute value of the correlation coefficient increases from left to right. In these sub-sample sets, the maximum value of r1 is 0.565, and the minimum value is -0.553. Considering that some sub-sample sets have a small number of samples and a large data dispersion, they were re-estimated, and their range fluctuated between -0.296 and 0.284. The maximum value of r2 is 0.510, and the minimum value is -0.696. Similarly, they were re-estimated, and the coefficients of the sub-sample sets with larger sample sizes fluctuated between -0.203 and 0.213. Overall, the sub-sample sets with negative correlations account for the vast majority, approximately accounting for 90\% of the total sample set. While the sub-sample sets with positive correlations are in the minority, possibly due to the large data dispersion of the sample data, thus resulting in a different correlation situation from the majority of the overall samples. Based on the correlation degree of the majority of samples, it can be inferred that the influence of the Eddington ratio on the light variation of quasars is more significant than redshift and the stationary wavelength, and is similar to the luminosity and black hole mass, having a relatively important modulation effect on the light variation process.

\begin{figure*} 
	\includegraphics[width=\textwidth]{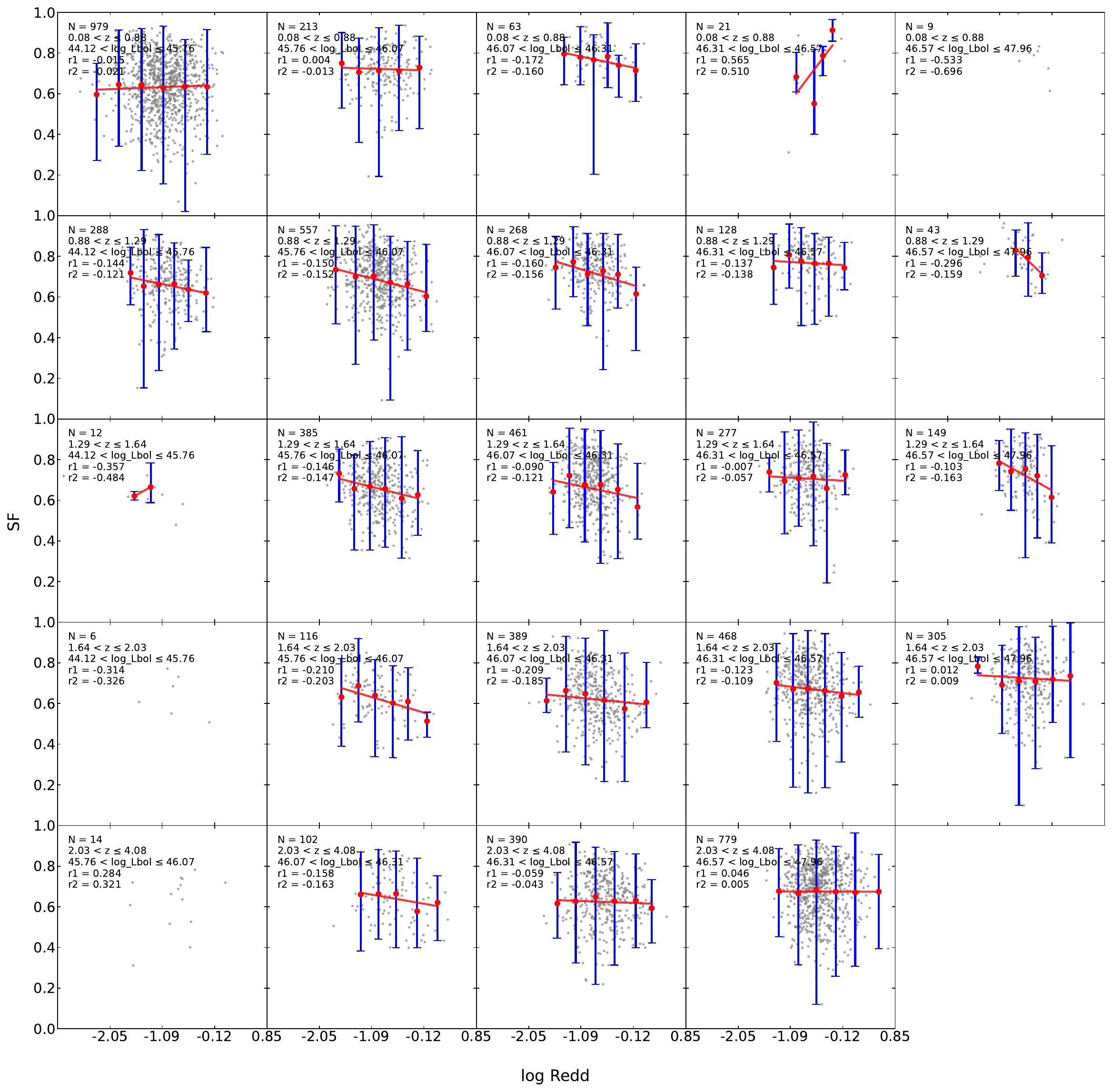}
    \caption{The dependence of the structural function on the Eddington ratio. In the figure, limits have been imposed on luminosity and redshift, and 24 sub-samples have been set up.}
    \label{fig:example_figure}
\end{figure*} 

According to Figure 8, the number of samples within each sub-sample set under the black hole mass bins does not vary significantly, indicating that the statistical results are relatively reliable. The distribution within the range of Eddington ratios from -2.05 to 0.85 is continuous, without any obvious interval breaks. This also indicates that the sample coverage in the Eddington ratio dimension is complete, and is consistent with the distribution characteristics of the 3-15 Eddington ratio frequency histogram. This ensures the continuity of the analysis of the evolution trend of the structural function with respect to the Eddington ratio. From the correlation coefficient, the correlations of the sub-sample sets show both positive and negative correlations, but overall they present a positive correlation. The subgraphs from left to right correspond to the redshift and black hole mass from low values to high values. It can be seen that in these sub-sample sets, the absolute value of the correlation coefficient from left to right increases. In these sub-sample sets, the maximum value of r1 is 0.354, and the minimum value is -0.010; the maximum value of r2 is 0.382, and the minimum value is -0.049. Overall, the sub-sample sets with positive correlations account for the vast majority, approximately accounting for 95\% of the total sample set. While the sub-sample sets with negative correlations are in the minority, possibly due to the large degree of dispersion of the sample data, thus resulting in a different correlation situation from the majority of the overall samples. Based on the correlation degree of the majority of samples, it can be inferred that the influence of the Eddington ratio on the light variation of quasars is more intense than that of redshift and the stationary wavelength, and is similar to the luminosity and black hole mass, having a more important modulation effect on the process of variability.

\begin{figure*} 
	\includegraphics[width=\textwidth]{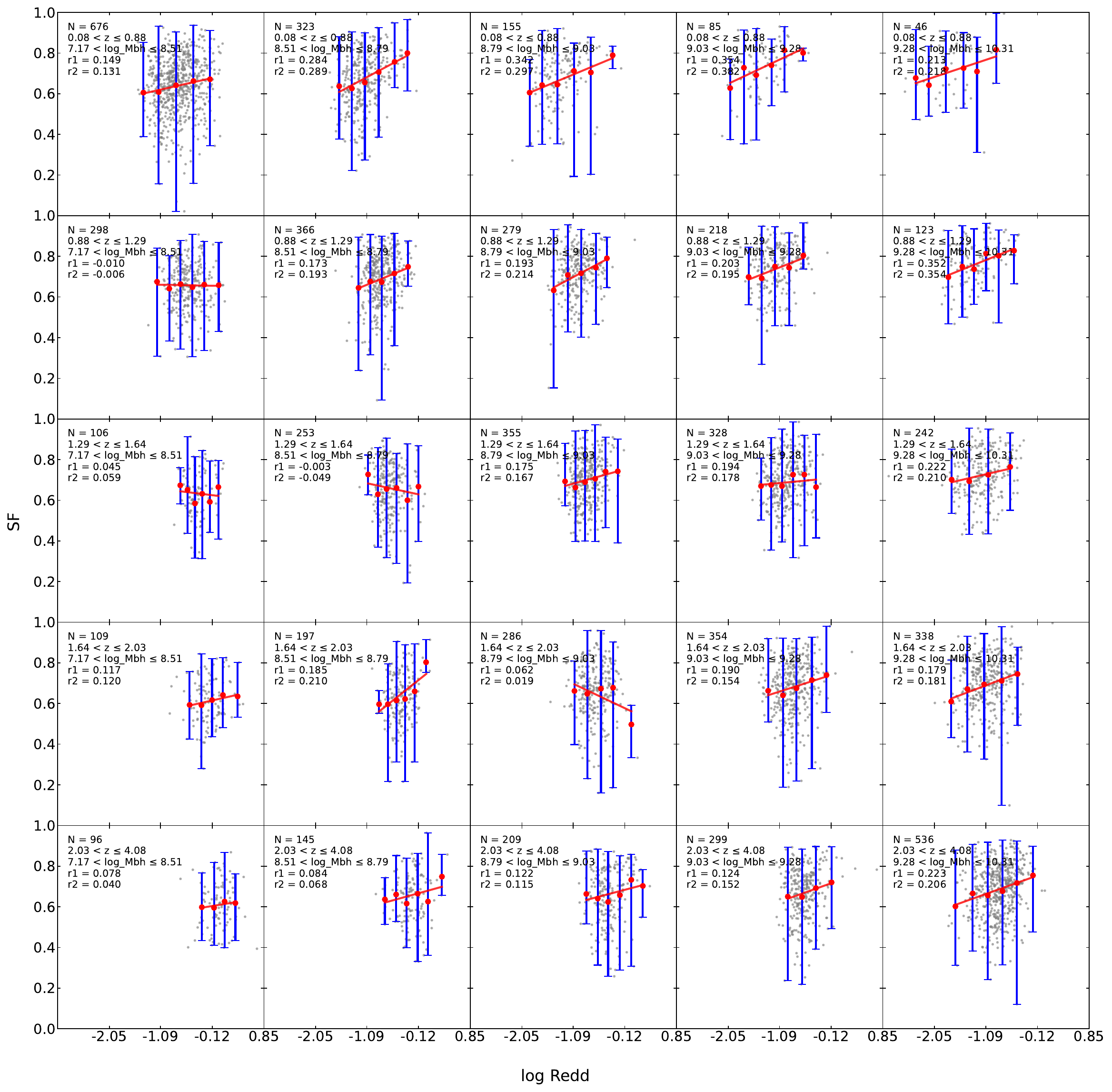}
    \caption{The dependence of the structure function on the Eddington ratio. In the figure, limits are imposed on the mass of the black hole and the redshift, and 25 sub-samples are set.}
    \label{fig:example_figure}
\end{figure*} 

\subsection{The dependence of the structure function on the accretion rate}

In the study of the structure function and accretion rate of quasars, the sub-sample sets were binned based on the redshift and the range of the stationary wavelength. Similarly, five sub-sample sets were divided. The dotted lines and annotations in the figure are the same as those in Figure 8.

From the sample size shown in the subgraph of Figure 9, it can be seen that the sample size is sufficient and the distribution is balanced. This is very helpful in avoiding statistical deviations caused by differences in sample size, and also provides an ideal statistical basis for the subsequent group comparison analysis. The overall distribution of accretion rate is continuous, ensuring the complete coverage of the accretion rate dimension for each sub-sample set. The slope of the fitting line between the structural function and the accretion rate in the sub-sample sets is positive, indicating a positive correlation between the quasar structural function and the accretion rate. Compared to the static wavelength and redshift, the degree of correlation is relatively significant, and the degree of significance is comparable to the black hole mass and Eddington ratio. From left to right in the subgraph, the range of redshift parameters becomes larger, and the correlation coefficients vary under these different redshift groups. The mean correlation coefficient in the low redshift bins shows a steeper upward trend, while in the high redshift bins it is relatively flat. This indicates that the structural function of mid-low redshift quasars is more sensitive to the changes in the accretion rate. In the sub-sample sets, the minimum value of r1 is 0.228 and the maximum value is 0.316; the minimum value of r2 is 0.224 and the maximum value is 0.316. All subgraphs show a positive correlation, and the differences in the correlation coefficients under different redshift groups indicate a weak reverse synergistic modulation effect of redshift on the dependence of the structural function on the accretion rate. It can be seen that the role of the accretion rate in the variability of quasars is extremely important.

\begin{figure*} 
	\includegraphics[width=\textwidth]{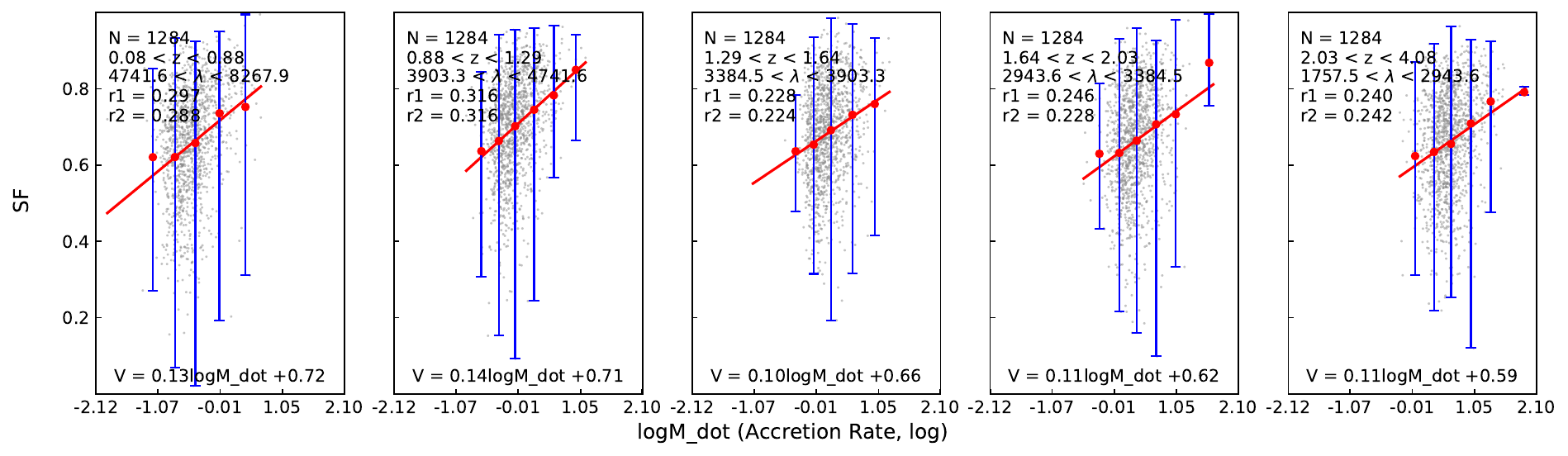}
    \caption{The dependence of the structure function on the accretion rate. In the figure, the rest wavelength and redshift are restricted, and five sub-sample sets are set.}
    \label{fig:example_figure}
\end{figure*} 

The analysis shows that the lower the Eddington ratio, the more obvious the positive correlation tendency of the quasar structure function with the increase in accretion rate; the higher the Eddington ratio, the positive correlation tendency between the structure function and the accretion rate is slightly weaker. Overall, the positive correlation presented, combined with the positive correlation between the structure function and the luminosity, as well as the positive correlation between the structure function and the black hole mass, suggests similarities in the accretion process of quasars with high luminosity, high black hole mass, and low Eddington ratio.

Based on the comprehensive analysis of parameter dependence, it was found that the correlation between the structure function and redshift and rest wavelength is relatively weak, indicating that the influence of redshift and rest wavelength on the variability of quasars is relatively small. Moreover, during the action of these two parameters, the Eddington ratio plays a modulation role. The dependence presented by quasars with low Eddington ratios is different from that of quasars with high Eddington ratios.

The parameters with a relatively high degree of correlation are luminosity, black hole mass, Eddington ratio, and accretion rate. For the two parameters of luminosity, the correlation with the structure function is positive, and the overall degree of correlation is relatively high; for the two parameters of black hole mass, the correlation with the structure function is also positive, and the degree of correlation is relatively large; the dependence on the previous two parameters is different. The correlation between the two parameters of Eddington ratio varies in different luminosity bins. In the luminosity bin, the correlation presented is negative, while in the black hole mass bin, the correlation is positive, but the positive and negative correlation degrees are comparable. For the accretion rate calculated from the parameters, the structure function has a positive correlation with it, and the Eddington ratio has a certain modulation effect on its dependence.

After comparison, it was found that there is a certain degree of dependence between the structure function and these parameters. However, black hole mass, Eddington ratio, and luminosity have the greatest influence on them. This further demonstrates the modulation effect of the intrinsic parameters of quasars in the variability, and also confirms that the variability evolution reflected by the structure function is a complex process of multi-parameter coupling. The modulation directions and intensities of different parameters are different. During these modulation processes, the synergistic modulation effect between black hole mass, Eddington ratio, and luminosity has a more significant impact on the variability. The accretion rate, as a key factor affecting the accretion process at the center of quasars, also shows a relatively significant correlation. Thus, the accretion rate plays a crucial role in the variability of quasars. After comparing with previous works, we found that the results obtained from this study are consistent and also have differences.

\section{Conclusions}

Using the sample of 6571 quasars from SDSS-DR7 (the seventh data release of the Sloan Digital Sky Survey), the relationships between the structure function and the static wavelength, redshift and quasar parameters, black hole mass, luminosity and Eddington ratio were studied. SDSS-DR7, as an important source of astronomical survey data, its quasar sample has significant value in cosmology and quasar physics research. Many studies have conducted measurements and analyses of quasar-related parameters based on this data set \citep{Abazajian2009}. Since the quality of each data point of each quasar is relatively good and the sample is large, we calculated the variability index of each quasar based on its long-term light curve, and by dividing the parameter space into small blocks, we unraveled the dependencies between different parameters. This research method based on a large sample and long-term light curve is consistent with the mainstream thinking of quasar variability studies in time-domain astronomy in recent years. Our main conclusions are as follows.

There is an extremely weak negative correlation between the structure function and redshift, and the correlation with the stationary wavelength has no unified direction. However, the correlation is not significant. This indicates that during the process of variability, the redshift and the stationary wavelength have very little influence on the central accretion process of quasars. Redshift, as an important indicator of the distance of quasars and the evolution of the universe, has always been a research hotspot regarding its influence on the variability of quasars. Previous studies have shown that the modulation effect of redshift on the variability of quasars is usually weaker than their intrinsic physical parameters \citep{Berk2004}. The modulation mechanism of the Eddington ratio reflected in these two dependencies indicates that the central accretion process of quasars will have a certain impact on the degree of variability, which is consistent with the common understanding that the accretion process is the core driving mechanism of quasars' variability \citep{Peterson1998}.

The correlation between the structural function and luminosity, Eddington ratio, and black hole mass is relatively strong. The correlations are positive, positive, negative, and positive respectively. The Eddington ratio shows a negative correlation with the structural function under the luminosity limit, and a positive correlation with the structural function under the black hole mass limit. However, the degree of correlation is not significantly different. This result indicates that the coupling mechanism of the synergy between luminosity and Eddington ratio, and between black hole mass and Eddington ratio is different. Luminosity and Eddington ratio may be direct factors causing the instability of the accretion disk of quasars, which is more consistent with the disk instability model in the mechanism of variability. And radiation pressure instability is one of the important causes of the instability of the accretion disk \citep{Lightman1974}. The black hole mass may affect the accretion disk through the depth of the gravitational potential well, causing changes in the accretion process of quasars, resulting in a more concentrated correlation of the structural function under the condition of a massive black hole. This is consistent with the theory that the supermassive black hole at the center of quasars dominates its radiation characteristics through the accretion process \citep{Kormendy2013}. In addition, the measurement of black hole mass in the SDSS-DR7 sample is often based on the Virial coefficient method of wide emission lines. The accuracy of this method has been verified and optimized through relevant studies.

Whether it is the weak correlation between the structure function and redshift, rest wavelength, or the stronger correlations between the structure function and luminosity, black hole mass, and Eddington ratio, all these demonstrate that the evolution process of the structure function of quasars is not caused by a single parameter, but rather the combined effect of multiple parameters. This also reflects that the reasons for the variability phenomenon of quasars are not unique, and there is a multi-factor coupling effect. This study involves samples in the optical band. It can be known that the variability of the optical band samples is mainly caused by the perturbation of the accretion disk, which is consistent with the theory that the optical radiation of quasars mainly comes from the accretion disk \citep{Krolik2000}; the luminosity, black hole mass, and Eddington ratio all represent the accretion process in the central region of quasars. Therefore, changes in the accretion process lead to the instability of the accretion disk, thereby triggering the variability and color variation of quasars over long or short periods of time, which is consistent with the research conclusion that large-amplitude quasar variability are caused by the instability of the accretion flow.

From the perspective of sample classification, the samples are generally divided into two major categories: the first category consists of high-redshift, high-luminosity, high-Eddington ratio, and massive black hole quasars, and the second category consists of low-redshift, low-luminosity, low-Eddington ratio, and low-mass black hole quasars. Based on the analysis of the structural functions and parameter dependencies in Chapter 3, it is found that the first type of quasars are in a state of intense accretion, with a relatively strong instability in the accretion disk of the quasars, and the structural function is more sensitive to changes in physical parameters. In this state, the degree of variability is more intense, which is consistent with the observed characteristics of high-luminosity quasars in the nuclear region with abundant fuel and intense accretion, and with significant amplitude of variability \citep{Jiang2020}; the second type of quasars is in a relatively gentle accretion state, with a relatively stable accretion disk, and the evolution of the structural function is more gradual. In this state, the degree of variability of the corresponding quasars is relatively slow, similar to the slowly varying variability characteristics of nearby low-luminosity quasar Spergel galaxies.

The influence of accretion rate on the variability of quasars is obvious. We found that the correlation presented in the case of high Eddington ratio is higher than that in the case of low Eddington ratio. According to the calculation formula of accretion rate, the difference in black hole mass and Eddington ratio will affect the accretion rate. Samples with low Eddington ratio correspond to high black hole mass, and it can be seen that both of them play an important role in the variability phenomenon of quasars. Supermassive black holes accelerate the accretion process in the central region, thereby affecting the accretion rate and causing drastic changes in the luminosity of quasars. This is consistent with the core theory that the accretion activity of supermassive black holes dominates the radiation intensity and variability characteristics of quasars \citep{Liu2024}, and also corresponds to the research result that the virial coefficient in the SDSS-DR7 sample is positively correlated with the dimensionless accretion rate of the black hole.

Through multi-dimensional statistical analysis and multi-parameter dependency research, the evolution law of the structure function of quasars was revealed, and the dominant factors of quasar variability were also revealed. The accretion process, as the core process affecting quasar light variation, contains the physical meaning embodied by the intrinsic parameters of quasars, the changing forms of quasar structure at different times, and also reflects the complexity and diversity of the process of quasar variability. This is consistent with the existing research cognition of the diversity and complexity of the mechanism of quasar variability \citep{Dong2018}.

\section*{Acknowledgements}

We are grateful for the work of the SDSS survey telescope staff. Their work provided data support for our research progress. At the same time, we would like to express our gratitude for the financial support provided by the Shaanxi Natural Science Foundation.

\section*{Data Availability}

Funding for the SDSS and SDSS-II has been provided by the Alfred P. Sloan Foundation, the Participating Institutions, the National Science Foundation, the U.S. Department of Energy, the National Aeronautics and Space Administration, the Japanese Monbukagakusho, the Max Planck Society, and the Higher Education Funding Council for England. The SDSS Web site is http://www.sdss.org/.



\bibliographystyle{mnras}
\bibliography{example} 





\bsp	
\label{lastpage}
\end{document}